# Super-resolution three-dimensional fluorescence and optical diffraction tomography of live cells using structured illumination generated by a digital micromirror device

**Seungwoo Shin,**[1,2] **Doyeon Kim,**[1,2,3] **Kyoohyun Kim,**[1,2,4] **and YongKeun Park**[1,2,3,*]

[1]*Department of Physics, Korea Advanced Institute of Science and Technology (KAIST), 291 Daehak-ro, Yuseong-Gu, Daejeon 34141, Republic of Korea*
[2]*KAIST Institute for Health Science and Technology, KAIST, Daejeon 34141, Republic of Korea,*
[3]*Tomocube, Inc., 48, Yuseong-daero 1184beon-gil, Yuseong-Gu, Daejeon 34051, Republic of Korea,*
[4]*Current affiliation: Biotechnology Center, Technische Universität Dresden, 01307 Dresden, Germany.*
[*]*yk.park@kaist.ac.kr*

**Abstract:** We present a multimodal approach for measuring the three-dimensional (3D) refractive index (RI) and fluorescence distributions of live cells by combining optical diffraction tomography (ODT) and 3D structured illumination microscopy (SIM). A digital micromirror device is utilized to generate structured illumination patterns for both ODT and SIM, which enables fast and stable measurements. To verify its feasibility and applicability, the proposed method is used to measure the 3D RI distribution and 3D fluorescence image of various samples, including a cluster of fluorescent beads, and the time-lapse 3D RI dynamics of fluorescent beads inside a HeLa cell, from which the trajectory of the beads in the HeLa cell is analyzed using spatiotemporal correlations.

## 1. Introduction

Three-dimensional (3D) imaging of live cells and their subcellular structures is crucial for the study of cell biology and provides invaluable information about related diseases [1]. Various 3D imaging approaches that were developed to investigate subcellular structures and constituents of live cells have several advantages, including high spatial and temporal resolution, molecular specificity, and noninvasiveness [2-5]. Fluorescence microscopy, a widely used microscopic method, can obtain molecule-specific information by labeling specific molecules inside cells with fluorescent proteins or dyes [2]. In particular, several super-resolution microscopic techniques that allow fluorescence imaging beyond the diffraction limit have been reported [6-10]. Among them, 3D structured illumination microscopy (SIM) provides 3D fluorescence images beyond both the lateral and the axial diffraction limit of conventional fluorescence microscopy by exploiting spatially structured excitation [10-13]. However, despite outstanding spatial resolution, 3D SIM has poor temporal resolution due to the time-consuming projection of the structured illumination sequence, mechanical axial scanning, and long acquisition time for dim fluorescence signals. Furthermore, 3D SIM may induce phototoxicity that could affect the cells, an inevitable problem of fluorescence microscopy [2].

In parallel, optical diffraction tomography (ODT) or 3D quantitative phase imaging (QPI) techniques have emerged as methods for label-free quantitative imaging of 3D refractive index (RI) distributions of biological samples [14-18]. To reconstruct 3D RI distribution, multiple two-dimensional (2D) holograms of a sample are measured using various incident angles. From the holograms, a 3D RI tomogram can be reconstructed via the principle of inverse light scattering [19]. Because the imaging contrast of ODT is a function of the RI, ODT is label-free and noninvasive, requires no preparation, has rapid image acquisition, and provides quantitative imaging [20, 21]. The RI distribution of live cells has provided structural and biochemical information about biological samples in studies in the fields of cell biology [22, 23], microbiology [24], hematology [25], infectious diseases [26], and biophysics [27]. Nonetheless, the RI generally provides limited molecular specificity [28], except for certain materials with distinct RI values such as lipids [29, 30] and metallic particles [31, 32].

Because both fluorescence imaging and QPI provide complementary information about cell pathophysiology, efforts have been made to combine and utilize both modalities for correlative bioimaging [33-35]. Several techniques for performing both 3D fluorescence imaging and 3D RI tomography of cells have been reported [30, 36-38]. Chowdhury et al. performed 3D SIM and ODT simultaneously utilizing a liquid crystal spatial light modulator (SLM) [38]. However, use of an SLM prevents fast imaging, one of the advantages of ODT. Furthermore, photobleaching and phototoxicity inevitably follow the use of an SLM.

In this paper, we propose and demonstrate with experiments a multimodal approach for measuring both the 3D RI and fluorescence distributions of live cells that combines ODT and 3D SIM using a digital micromirror device (DMD). Time-multiplexed structured illumination patterns displayed by the DMD are used to perform 3D super-resolution fluorescence imaging and obtain 3D RI tomograms of samples simultaneously and effectively. The fast dynamic response of the DMD enables imaging of live cells with high spatiotemporal resolution, molecular specificity, and reduced damage from phototoxicity. Because the same optical imaging system with the DMD is used for both modes, the correlation between 3D fluorescence and 3D RI data can be easily analyzed. The present method will open a new avenue into the study of cell biology, medical research, and diagnosis.

## 2. Methods

### 2.1 Optical setup

Figure 1 shows the schematic of the optical setup, equipped with a DMD (DLP6500EVM, Texas Instruments, Dallas, TX, USA), used in our experiments. This setup was able to operate in both the ODT mode and the 3D SIM mode [Figs. 1(a)−(c)] by modifying a commercial ODT setup (HT-1H, Tomocube Inc., Daejeon, South Korea) to incorporate fluorescence excitation.

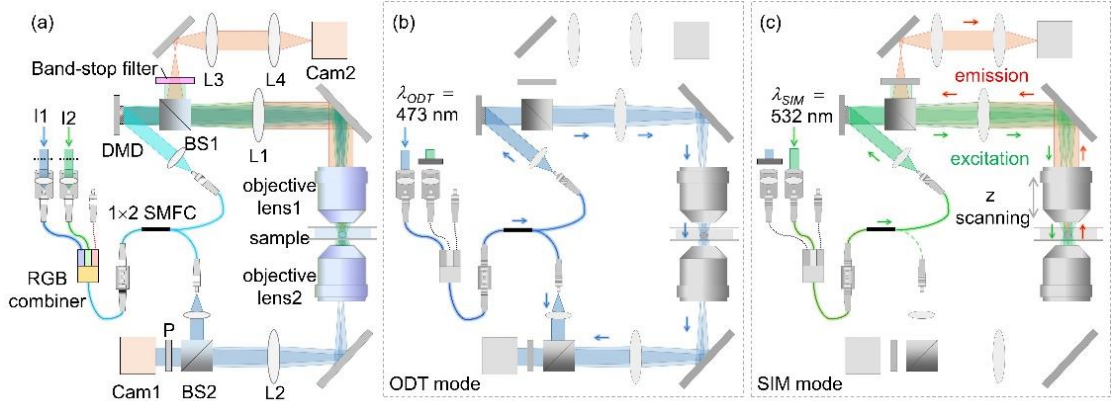

Fig. 1. (a) Experimental setup: I1, I2, irises; 1×2 SMFC, 1×2 single-mode fiber optic coupler; BS1, BS2, beam splitters; L1-L4, lenses ($f_1$ = 250 mm, $f_2$ = 180 mm, $f_3$ = $f_4$ = 150 mm); P, polarizer; Cam1, Cam2, cameras. (b) Setup for ODT based on a Mach-Zehnder interferometer equipped with a DMD. (c) Setup for 3D SIM based on epifluorescence microscopy.

To minimize photobleaching and phototoxicity that occur in fluorescence imaging of live cells, we used two lasers with different wavelengths [$\lambda_{ODT}$ = 473 nm (MSL-FN-473, CNI Laser, Changchun, P.R. China) and $\lambda_{SIM}$ = 532 nm (LSS-0532, Laserglow Technologies, Toronto, ON, Canada)] for coherent imaging and fluorescence excitation, respectively. To simplify the setup, the two laser beams were coupled into a single-mode fiber using an RGB combiner (RGB30HF, Thorlabs, Inc., Newton, NJ, USA). By selectively blocking one of the laser beams, the present setup alternated between the ODT mode and the 3D SIM mode. In both modes, the illumination part that included the DMD, lens L1 ($f_1$ = 250 mm), and objective lens 1 [UPLSAPO 60XW, numerical aperture (NA) 1.2, Olympus Inc., Tokyo, Japan] was the same, but the modes used different measurement schemes [Figs. 1(b) and (c)].

In the ODT mode, a sample is illuminated with various illumination patterns and the optical fields diffracted from the sample are measured using Mach-Zehnder interferometry [Fig. 1(b)]. A 1 × 2 single-mode fiber optic coupler (SMFC) (TW560R3F1, Thorlabs) splits the laser beam ($\lambda_{ODT}$ = 473 nm) into a reference beam and a sample beam. To control the angle of the beam impinging on a sample, the DMD displays time-multiplexed patterns that are projected onto the sample [39]. Using objective lens 2 (UPLSAPO 60XW, NA 1.2) and lens L2, ($f_2$ = 180 mm), the beam diffracted from the sample is collected and projected onto the image plane, where it interferes with the slightly tilted reference beam to generate an off-axis hologram. Multiple 2D holograms with various illumination patterns are recorded using an image sensor (Cam1; FL3-U3-13Y3M-C, FLIR Systems, Wilsonville, OR, USA).

In the 3D SIM mode, the same DMD generates structured illumination patterns that excite fluorescent molecules in the sample plane [Fig. 1(c)]. An excitation laser beam ($\lambda_{SIM}$ = 532 nm) is reflected from the DMD displaying a time-multiplexed pattern, and then the time-multiplex pattern is projected onto a sample through lens L1 and objective lens 1. This excitation configuration enables fast and stable generation of structured illumination patterns. Based on an epifluorescence detection scheme, fluorescence images of the sample are obtained in sequence for each structured illumination pattern at each axial position using the same objective lens. An sCMOS camera (Cam2; C11440-22C, Hamamatsu Photonics, Hamamatsu City, Japan) records the fluorescence images via objective lens 1 and relay lenses L1, L3, and L4 ($f_3$ = $f_4$ = 150 mm). A beam splitter (BS1) and a band-stop filter (NF01-532U-25) are used to block excitation beams in the path of the fluorescence beam.

*2.2 Optical diffraction tomography*

To reconstruct the 3D RI distribution of a sample, a modified commercial ODT (HT-1H, Tomocube) was utilized in an optical setup based on Mach-Zehnder interferometry using a DMD (Fig. 2) [21, 39].

The DMD displays a time-multiplexed pattern [21] that systematically controls the illumination angles. The superposition of the two plane waves from the DMD results in the formation of structured illumination that impinges on the sample [Figs. 2(a) and (b)]. The two plane waves diffracted by the displayed pattern in the DMD are mutually conjugated optical fields. Thus, the phase shift $\Delta\phi$ between the plane waves and the incident angles of the plane waves on the sample can be precisely controlled by the pattern displayed by the DMD. In addition, the time-multiplexing method significantly reduces unwanted diffraction noise that usually occurs when binary control is used in a DMD. For efficiency, the limited number of time-multiplexed patterns stored on the DMD control board consists of a three-binary-pattern sequence.

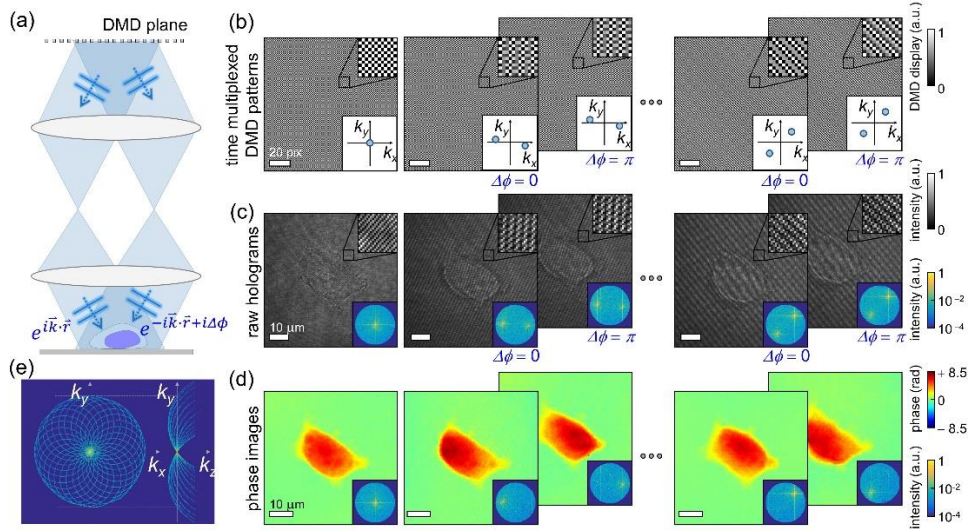

Fig. 2. Reconstruction of the 3D RI tomogram of a sample. (a) Structured illumination generated by a DMD is the superposition of two plane waves with a controlled wave vector and phase shift $\Delta\phi$. (b) Time-multiplexed patterns displayed on the DMD and (c) the corresponding holograms of a sample. Insets: magnified image (upper) and corresponding Fourier spectrum in the spatial frequency $k$ space (lower). Except for sample illumination at a normal angle, a pair of patterns projects two plane waves with different phase shifts onto a sample to decompose diffracted optical fields with respect to individual plane wave components. (d) Phase images of the retrieved optical fields corresponding to the plane wave components with various incident angles. Inset: corresponding Fourier spectrum. (e) A mapped object function in 3D Fourier space using the principle of ODT.

In addition to the normal angle, we used 28 time-multiplexed patterns (14 pairs) to generate 28 plane waves at different azimuthal angles so that the illumination beams scanned a circular pattern within the NA of the condenser lens [Fig. 2(b)]. To decompose the time-multiplex pattern into two plane waves, a pair of structured patterns with different relative phase shifts, 0 and π, respectively, was needed. Therefore, 28 time-multiplexed patterns were constructed to produce the 28 plane waves used for circular scanning.

The optical fields diffracted from a sample are retrieved from the measured holograms corresponding to the time-multiplexed structured illuminations [Fig. 2(c)] using a field retrieval algorithm based on Fourier transform [insets in Fig. 2(c)] [39]. Then, using the known phase shifts in the structured illuminations, the retrieved optical fields under the structured illuminations are decomposed into the scattered fields from the sample with respect to plane-wave illumination components as follows:

$$\begin{bmatrix} S_1 \\ S_2 \end{bmatrix} = \begin{bmatrix} 1 & e^{i\Delta\phi_1} \\ 1 & e^{i\Delta\phi_2} \end{bmatrix}^{-1} \begin{bmatrix} E_1 \\ E_2 \end{bmatrix}, \tag{1}$$

where $E_{1,2}$, $\Delta\phi_{1,2}$, and $S_{1,2}$ are the retrieved fields, the known phase shifts, and the decomposed scattered fields for each plane-wave component [Fig. 2(d)], respectively. Before the decomposition, each of the retrieved fields is normalized for one plane-wave component, in order to compensate the random fluctuations of the atmosphere and the mechanical instabilities of the interferometer. From the obtained scattered fields, an object function in 3D Fourier space is mapped using the principle of ODT to reconstruct the 3D RI distribution of the sample [Fig. 2(e)] [14, 19, 40]. To compensate for the uncollected side scattering signals that result from the limited NA of the objective lenses, the iterative regularization algorithm based on the assumption of non-negativity was used [41, 42].

*2.3 3D Structured illumination microscopy (SIM)*

To obtain a 3D fluorescence image with subdiffraction resolution, we used the principle of SIM (Fig. 3) [10, 11, 14]. SIM images were reconstructed from fluorescence images obtained using structured excitation patterns.

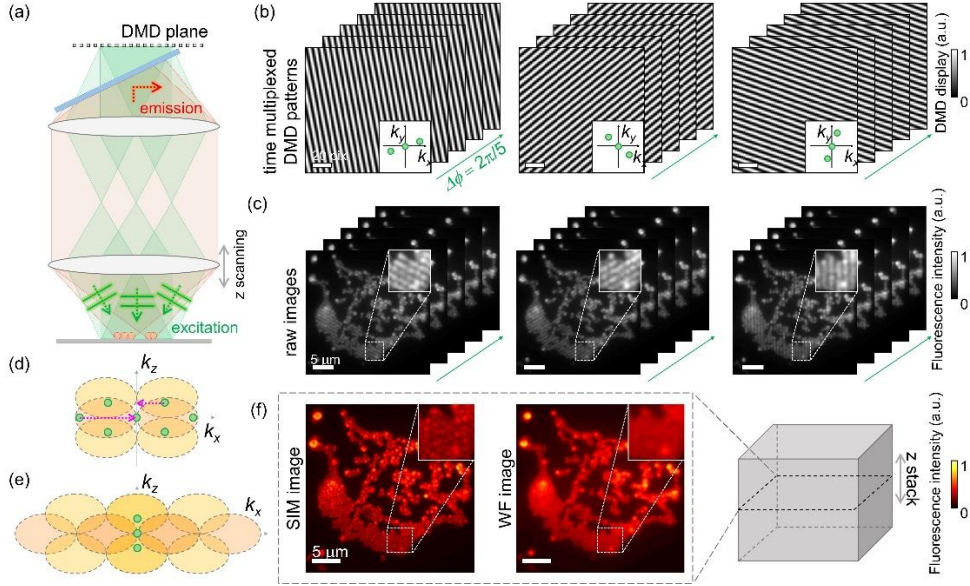

Fig. 3. Reconstruction of the 3D fluorescence image of a sample. (a) Epifluorescence microscopy geometry for structured illumination excitation of a sample and recording its fluorescence emission. A DMD controls both the phase shift and the illumination angles of three plane waves that result in the structured excitation pattern on a sample. (b) Time-multiplexed patterns displayed on the DMD (inset: corresponding Fourier spectrum) and (c) corresponding fluorescence images of a sample (inset: magnified images). (d) and (e) Illustrations in Fourier space for describing the SIM reconstruction from (d) measured fluorescence images to (e) reconstructed SIM images. (f) Reconstructed SIM and wide-field fluorescence (WF) images of a cluster of red fluorescent microspheres.

To project a structured excitation pattern onto a sample, the DMD is used to rapidly generate a time-multiplexed pattern comprising three plane waves [Figs. 3(a) and (b)]. The two oblique plane waves are mutually conjugated fields that are diffracted by the displayed pattern, while the third plane wave has a normal incident angle. The various patterns displayed on the DMD systematically control the phase shift and the incident angles of the two oblique plane waves relative to the residual plane wave. To reconstruct the SIM images, 15 time-multiplexed patterns are displayed as a combination of three azimuthal angles and five phase shifts [Fig. 3(b)] [10].

A 2D SIM image is reconstructed from the recorded fluorescence image [43]. Fluorescence emission results from an excitation intensity pattern. A recorded fluorescence image is the sum of component fluorescence images with the five different phase shifts, $F_n = \sum_{m=-2}^{2} C_m e^{m\, i\Delta\phi_n}$, where $n$ is the phase-shift index, $F_n$ is a recorded fluorescence image, $C_m$ is an indexed component image, and $\Delta\phi_n$ is the relative phase shift of the three plane waves that result in structured illumination. Using the known phase shifts controlled by the DMD, the component images can be decomposed as follows:

$$\begin{bmatrix} C_{-2} \\ C_{-1} \\ C_0 \\ C_1 \\ C_2 \end{bmatrix} = \begin{bmatrix} e^{-2i\Delta\phi_1} & e^{-i\Delta\phi_1} & 1 & e^{i\Delta\phi_1} & e^{2i\Delta\phi_1} \\ e^{-2i\Delta\phi_2} & e^{-i\Delta\phi_2} & 1 & e^{i\Delta\phi_2} & e^{2i\Delta\phi_2} \\ e^{-2i\Delta\phi_3} & e^{-i\Delta\phi_3} & 1 & e^{i\Delta\phi_3} & e^{2i\Delta\phi_3} \\ e^{-2i\Delta\phi_4} & e^{-i\Delta\phi_4} & 1 & e^{i\Delta\phi_4} & e^{2i\Delta\phi_4} \\ e^{-2i\Delta\phi_5} & e^{-i\Delta\phi_5} & 1 & e^{i\Delta\phi_5} & e^{2i\Delta\phi_5} \end{bmatrix}^{-1} \begin{bmatrix} F_1 \\ F_2 \\ F_3 \\ F_4 \\ F_5 \end{bmatrix}. \qquad (2)$$

Next, the Fourier spectra of the decomposed component images are shifted in the Fourier space to reconstruct a SIM image using the principle of SIM [Figs. 3(d) and (e)]. Furthermore, a 3D SIM image can be obtained by stacking 100 reconstructed SIM images using a 150-nm interval along the axial direction [Fig. 3(f)]. To compare the spatial resolution of the 3D SIM images with the 3D wide-field fluorescence (WF) images, 3D WF images are reconstructed by stacking the 2D WF images along the axial direction.

*2.4 Sample preparation*

We used red fluorescent polystyrene (PS) beads for the nonbiological sample and HeLa cells for the biological sample.

The red fluorescent PS beads (L3280, Sigma-Aldrich, St. Louis, MO, USA) were diluted in distilled water, placed on a coverslip, and dried overnight so that the beads adhered to the surface of the coverslip. Before measurements were taken, phosphate-buffered saline (PBS) solution (pH 7.4, 50 mM, Welgene, Gyeongsangbuk-do, South Korea) was gently dropped on the coverslip, which was then covered by another coverslip.

HeLa cells, a human cervical cancer cell line, were cultured in Dulbecco's modified Eagle's medium (DMEM) (Welgene) with 10% fetal bovine serum (FBS) (Welgene) and 1% penicillin streptomycin [100× (v/v), Welgene] inside a humidified incubator (37 °C, 5% $CO_2$, 95% air). After 12 h of incubation, trypsin EDTA (Welgene) was applied to detach the cells from the culture flask. The detached cells were then collected via centrifugation (1000 rpm for 3 min) and incubated on a coverslip using the same previous culture conditions for 3 h. The diluted red fluorescent PS beads (100× diluted from stock solution) were added to the culture medium and became inserted as fluorescent probes inside the HeLa cells. Cells on the coverslip were maintained with the bead solution for 12 h and then washed three times with PBS solution.

## 3. Results and discussion

*3.1 Multimodal imaging of a monolayer of fluorescent beads*

To demonstrate the validity of the proposed method, we measured the 3D RI distribution and 3D fluorescence image of a cluster of red fluorescent 500-nm-diameter PS beads (Fig. 4). In Figs. 4(a), (c), and (e) for the whole field of view (FOV) and in Figs. 4(b), (d), and (f) for the magnified FOV, a monolayer of fluorescent beads is clearly seen in both the 3D RI distribution and the 3D fluorescence image. In addition, the reconstructed 3D RI distribution matches the reconstructed 3D SIM and WF images well. This clearly shows the feasibility of our method.

The lateral and axial resolutions of ODT in the proposed method were 172 nm and 665 nm, respectively, as calculated from the maximum achievable spatial frequencies of the setup along the lateral and axial directions [11]. Because the axial resolution of ODT in the proposed method was larger than the diameter of the PS beads, the reconstructed RI distribution of the beads showed elongated images in the axial direction as well as RI values smaller than the known RI value of PS ($n$ = 1.608 at $\lambda$ = 473 nm).

The cross-sectional slices of the reconstructed 3D SIM images [Figs. 4(c) and (d)] show the ability to perform optical sectioning and enhanced lateral and axial resolutions of 3D SIM compared to 3D WF [Figs. 4(e) and (f)]. Moreover, the extended lateral and axial bandwidths are clearly seen in the 3D Fourier spectrum of the 3D SIM image in Fig. 4(h) compared to those seen in the spectrum of the 3D WF image [Fig. 4(g)]. In particular, the cross-sectional axial slice of the 3D Fourier spectrum of the 3D SIM does not exhibit the missing cone, which causes severe deterioration of the 3D WF image along the axial direction [10].

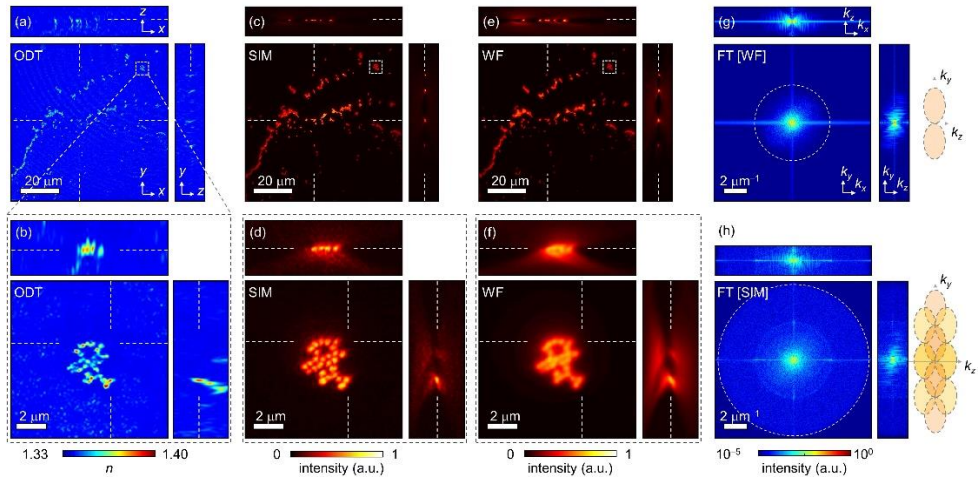

Fig. 4. 3D RI distribution and 3D fluorescence image of a cluster of red fluorescent PS beads. (a) and (b) Cross-sectional slices of reconstructed 3D RI and (c)-(f) 3D fluorescence distributions in *x-y*, *x-z*, and *y-z* planes of a cluster of red fluorescent 500-nm-diameter PS beads. The slices in each column were reconstructed using (a) and (b) ODT, (c) and (d) SIM, and (e) and (f) WF. The slices in (b), (d), and (f) are magnified images of the area in the dashed boxes in (a), (c), and (e). (g) and (h) Cross-sectional slices of the 3D Fourier spectrum of 3D fluorescence distributions reconstructed using WF and SIM, respectively. Illustrations on the right-hand side of (g) and (h) are representations of the theoretical transfer function of WF and SIM.

*3.2 4D Time-lapse imaging of fluorescent beads inside a HeLa cell*

The proposed multimodal method combines the advantages of ODT and 3D SIM by providing label-free 3D RI distribution with superior temporal resolution along with high molecular specificity. To demonstrate the synergistic advantages and the applicability of the proposed method, we measured the dynamics of the 3D RI distribution of 500-nm-diameter fluorescent PS beads ingested by a HeLa cell (Fig. 5).

First we measured the 3D RI distribution and the 3D fluorescence image of PS beads inside a HeLa cell using the ODT and 3D SIM modes, respectively [Figs. 5(a) and (b)]. The 3D isosurface of the cell [Fig. 5(c)] rendered from the reconstructed 3D RI distribution [Fig. 5(a)] shows that the fluorescent beads inside the cell are difficult to distinguish from the surrounding cytoplasm because the axial resolution of ODT is larger than the diameter of the beads (see Section 3.1). In contrast, the correlation of the 3D RI distribution with the measured 3D SIM image clearly shows the fluorescent beads [Fig. 5(d)], where the reconstructed RI values of the identified fluorescent beads replaced the known RI value of PS ($n$ = 1.608 at $\lambda$ = 473 nm).

After identifying the fluorescent beads in the reconstructed 3D RI distribution, we obtained time-lapse 3D RI distribution measurements of the HeLa cell every 30 s over 1 h using ODT [Fig. 5(e)]. Because the subcellular dynamics of the HeLa cell was slow compared to the tomographic acquisition speed, the reconstructed 3D RI distributions of the beads between consecutive measurements were highly correlated, which enabled the 3D dynamics of the fluorescent beads inside the HeLa cell to be tracked without additional 3D SIM measurements [Fig. 5(e)].

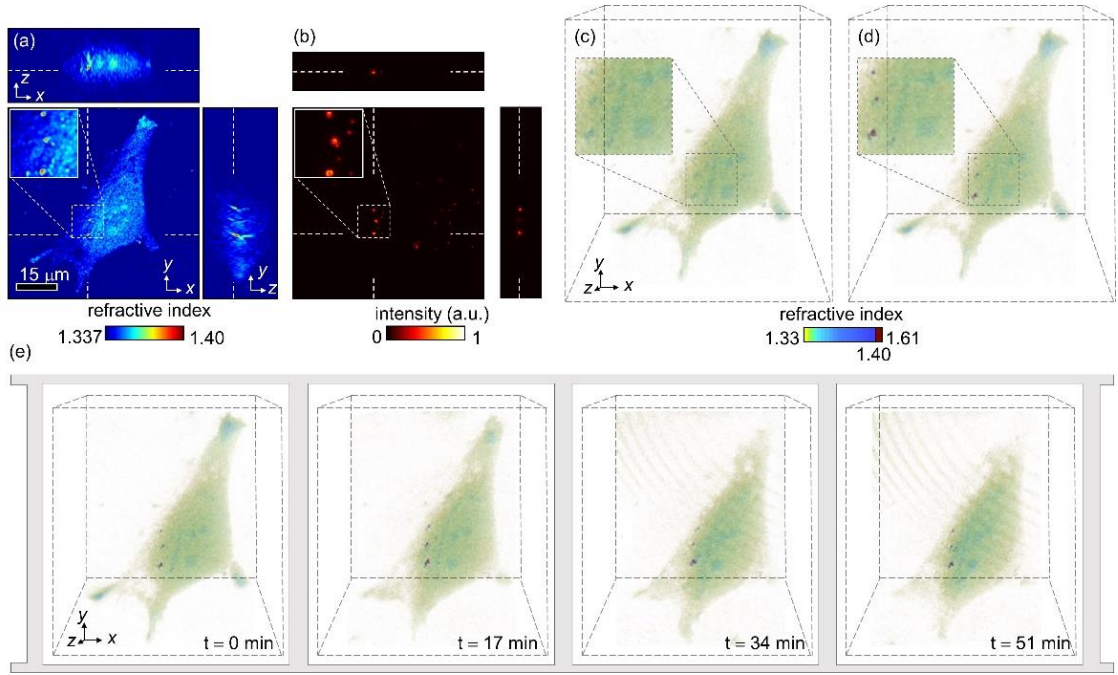

Fig. 5. Demonstration of the applicability of the proposed method. Cross-sectional slices of (a) the reconstructed 3D RI and (b) fluorescence distributions in the *x-y*, *x-z*, and *y-z* planes of a HeLa cell containing fluorescent PS beads. (c) 3D rendered isosurfaces of the HeLa cell from the reconstructed 3D RI distributions in (a). (d) 3D rendered isosurfaces of the HeLa cell after labeling the fluorescent beads by combining the reconstructed 3D RI and fluorescence distributions in (a) and (b). (e) Time-lapse 3D RI distributions of the HeLa cell. The labeled fluorescent beads are tracked using spatiotemporal correlations (see Visualization 1 and Visualization 2).

## 4.  Conclusion

In this paper, we presented our proposed multimodal approach to imaging that combines ODT and 3D SIM using a DMD and experimentally demonstrated its feasibility and applicability. Each mode can be operated separately by using different illumination wavelengths. As a result, the combination of the respective advantages of ODT and 3D SIM yields the synergistic effects of our method. To demonstrate the feasibility of the proposed method, we reconstructed 3D RI and fluorescence distributions of a planar cluster of fluorescent PS beads using the ODT and 3D SIM modes, respectively. Furthermore, to demonstrate the applicability of our method, we reconstructed 3D fluorescence and time-lapse 3D RI distributions of beads inside a HeLa cell. By exploiting the spatiotemporal correlations of 3D fluorescence and time-lapse 3D RI distributions, the beads were tracked as the cell changed morphologically.

Our proposed method provides 3D structural and biochemical information on fluorescent-labeled live cells with high molecular specificity. The advantages of our method, such as its high spatiotemporal resolution and minimization of the photobleaching problem, provide it the potential for use in a wide range of applications, including the study of subcellular dynamics over a long time.

## Acknowledgements

This work was supported by KAIST, BK21+ program, Tomocube Inc., and National Research Foundation of Korea (2017M3C1A3013923, 2015R1A3A2066550, 2014K1A3A1A09063027). All the authors have financial interests in Tomocube Inc., a company that commercializes ODT and QPI instruments and is one of the sponsors of the work.